\documentclass[useAMS]{mn2e}

\usepackage{graphicx}

\title[Lyman $\alpha$ emission from Damped Lyman $\alpha$ Systems]{Faint Resonantly 
Scattered  Lyman $\alpha$ Emission from  the Absorption Troughs
of Damped Lyman $\alpha$ Systems at $z\sim 3$}
\author[Michael Rauch \& Martin Haehnelt]{Michael Rauch,$^{1}$\thanks{E-mail:
mr@obs.carnegiescience.edu} and  Martin G. Haehnelt,$^{2}$\thanks{E-mail: haehnelt@ast.cam.ac.uk}\\
$^{1}$Carnegie Observatories, 813 Santa Barbara Street, Pasadena, CA 91101, USA\\
$^{2}$Kavli Institute for Cosmology and Institute of Astronomy, Cambridge University, Madingley Road,  Cambridge CB3 0HA, UK}
\begin{document}

%\date{Research Note; submitted to astro-ph only (using MNRAS stationery)}

\pagerange{\pageref{firstpage}--\pageref{lastpage}} \pubyear{2010}

\maketitle

\label{firstpage}

\begin{abstract}

We demonstrate that the  Ly$\alpha$ emission 
in the absorption troughs of a large sample of stacked 
damped Ly$\alpha$ absorption systems  (DLAS) presented  by  Rahmani et al (2010) is consistent 
with the spectral profiles and luminosities of
a recently detected population  of faint Ly$\alpha$ emitters at $z\sim 3$.
This result supports the suggestion that the faint emitters  are  to be identified 
with the host galaxies of DLAS at these redshifts.

\end{abstract}


\section{Introduction}

An  ultra-deep spectroscopic blind survey 
(Rauch et al 2008; hereafter R08) has  revealed the existence of
numerous, faint,
spatially extended   Ly$\alpha$ emitters at $z \sim 3$. 
The inferred rate of incidence of the emitters, and the observed signatures of resonant radiative transfer  
suggest that the emission originates from regions corresponding to  high column density HI QSO absorption systems.
R08  proposed  that these objects may thus be identified with the elusive host galaxies of DLAS and Lyman Limit systems (LLS). Their study  showed  that even objects with the low luminosities and low masses characteristic of high redshift dwarf galaxies
are surrounded by extended gaseous halos, whose intrinsic Ly$\alpha$ emission can typically
be traced out to radii of at least 4".
\smallskip

Rahmani et al (2010) recently published a high signal-to-noise
composite spectrum of the absorption trough of high redshift DLAS from the
Sloan Digital Sky Survey (SDSS), to search for Ly$\alpha$ emission from the underlying galaxies
giving rise to the DLAS. Their sample represents low resolution (R=2000)
spectra of 341 DLAS, with column densities exceeding $\log$N$_{\rm HI}$=20.62.
The DLAS where shifted to the rest frame and then averaged according
to various prescriptions.  The individual QSO
spectra used represent the integrated light  of fibers with 3" diameter. Since many 
of the R08 objects have relatively compact cores of Ly$\alpha$ emission  (on top
of more extended low light level emission),
a significant fraction of the Ly$\alpha$ luminosity of such galaxies should occur close enough
to the line-of-sight  to be recorded within the fiber radius,
{\it assuming that these emitters
are in fact associated with DLAS}. 
Rahmani et al establish  upper limits on the flux in a central wavelength region of the stacked absorption troughs.
Based on this result they conclude that the low mean flux permitted by their analysis contradicts, at the $3-\sigma$ level, the suggestion 
that DLAS could correspond to 
a population of Ly$\alpha$ emitters with a mean flux as high as found by R08.
We argue here that this conclusion is based on an inappropriate model
of the emission profile used by Rahmani et al.
When allowance is made for the resonant scattering of the Ly$\alpha$
photons with their inherent wavelength shifts and asymmetries, 
their combined DLA profile is consistent with substantial Ly$\alpha$ flux, 
and supports, rather than contradicts, the correspondence between DLAS 
and Ly$\alpha$ emitters proposed by R08.

\section[]{The expected  Ly$\alpha$ emission profile}

\begin{figure}
\includegraphics[scale=.5,angle=0,keepaspectratio = true]{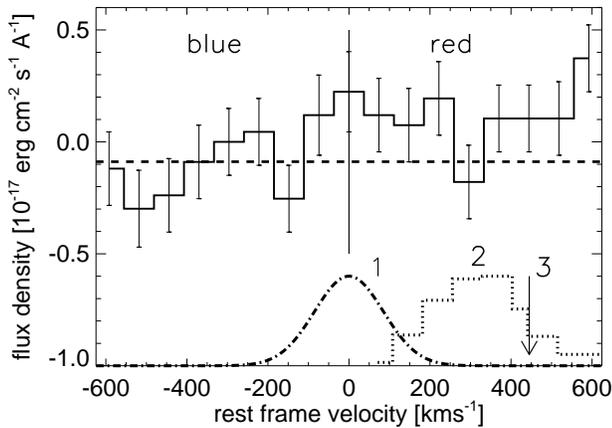}
\caption{Residuals of the flux density at the bottom of the stacked DLA
trough published by Rahmani et al 2010,  obtained by subtracting
the DLA model in their figure 1 (weighted mean with 5\% clipping, their rightmost panel) from the flux histogram. The vertical
line is centered on the systemic velocity of the DLAS absorption line.
The horizontal dashed line shows the mean flux in the left (blue) half of the
absorption line, as a better estimate of the zero level. 
The bottom of the plot (arbitrary offset with arbitrary units) shows: (1) the shape of the selection window of Rahmani et al (a Gaussian with FWHM 200 kms$^{-1}$); (2) for illustrative purposes,
a theoretical emission profile for a spherical DLA halo expanding homologously with up to 200 kms$^{-1}$, with column density $2\times10^{20}$cm$^{-2}$ and temperature T=$2\times10^4$ K (Zheng \& Miralda-Escud\'e 2002);
(3) the mean position (arrow) of the Ly$\alpha$ emission line in the Lyman break galaxy sample by Steidel
et al 2010 (the velocities extend redward beyond the frame of the plot).
\label{fig1}}
\end{figure}

Rahmani et al searched for Ly$\alpha$ emission at the center of the DLA line, assuming a Gaussian velocity distribution with a width of 200kms$^{-1}$. This model would be appropriate for optically thin emission,
broadened only by galactic velocity dispersion and the instrumental profile. However, numerous theoretical studies of radiative transfer through the very
optically thick DLAS (e.g., Zheng \& Miralda-Escud\'e 2002; Dijkstra et al 2006; Verhamme et al 2006;  and references therein) have shown that the generic emission line profile is double-humped, with apparent velocity shifts
of hundreds of kms$^{-1}$ between the blue and red components, and little emission at all at the systemic redshift. The profile is further modified
by the motion of the gas. 
Observationally,  in high redshift galaxies the red peak dominates, and the blue peak is highly reduced in size. This pattern
can be accomplished by Ly$\alpha$ propagating through an expanding halo, partial absorption by intervening Ly$\alpha$ forest clouds, or a combination of 
both (e.g., Dijkstra et al 2006; Barnes \& Haehnelt 2010; Laursen et al. 2010). The pattern of a dominant red peak, absorption trough, and faint blue peak is clearly seen in the brighter of the sources found  by R08, and a similar pattern (with even less blue emission) is seen
in the much brighter Lyman break galaxies. In the latter objects the red line suffers a mean shift
of 445 kms$^{-1}$ relative to the systemic velocity (Steidel et al 2010).   Similarly large shifts are predicted  for Ly$\alpha$ photons passing through simple static slabs of HI gas with column densities typical of DLAS (e.g., Dijkstra et al 2006; Hansen \& Oh 2006). 
Thus, unless there are many objects where the Ly$\alpha$ escapes 
through optically thin holes in the HI distribution (as  may be produced through photoionization by an AGN), there should be little
flux at the center of most DLA troughs. Note that the population of Ly$\alpha$ emitters, at least at its bright end, where there are more detailed
observations,  harbors only a small contamination by AGN (e.g. Ouchi et al 2008). 

Among the non-AGN Ly$\alpha$ emitters the red peak can occur over a wide range of velocities on the  red (right hand) side of the absorption trough, with the position determined
by column density, temperature, kinematics and possibly dust in a not necessarily trivial
way. The intensity in the blue half of the trough should generally be weaker in proportion to the ratio between
blue and red peaks. Thus we expect an upward jump in the flux density 
when going from the blue to the red half of the bottom of the DLA line. Such a spectral distribution
of the  Ly$\alpha$ emission  is indeed suggested  by the  appearance of the stacked spectrum of Rahmani et al.  In our Fig.~1  we show the residuals obtained after we  subtracted their model of the DLA trough  
from  the  flux density in the  rightmost panel (weighted mean with 5\% clipping) of their figure 1. We also compare the spectral profile 
assumed by Rahmani et al. with a more realistic, predicted profile of an individual emitter  taking into account 
the expected resonant  scattering of the Ly$\alpha$ emission.  We further show
the mean observed position of the Ly$\alpha$ line in a sample of Lyman break galaxies. Note that for a large  ensemble 
of emitters a broad distribution of the location and width of the "red peak"  would be expected. 
For a stacked spectrum like the one of Rahmani et al. this  should  lead to a  broad emission feature  on the red side of the trough, extended over several hundred km/s.  Unfortunately, because of their faintness, we cannot determine the systemic  redshifts of the R08 emitters so it is not
possible to stack the R08 spectra to directly determine the shape of
the combined emission feature.

The mean observed flux density over the entire trough in Fig.~1 is $(3.4\pm7.8)\times10^{-19}$ erg s$^{-1}$cm$^{-2}$\AA, i.e., consistent with zero, in agreement with Rahmani et al.'s  
statement that a global correction of the zero level has been performed for each spectrum to account for imperfections in the sky background subtraction. However, 
the Kolmogorov-Smirnov probability for the residuals in the 16 pixels with error bars in their figure to be consistent with zero flux throughout is only 1\%. The probability
for the flux density in the blue half and the red half of the DLA trough (the bluest 8 pixels and the reddest 8 pixels of the residuals) to be
drawn from the same sample is likewise less than 1\%, i.e., the flux
levels in the blue and red half of the trough are significantly different,
and not consistent with being at the true zero level.
The mean flux density in the left side of the trough (bluest 8 pixels plus half of the central pixel) and the right side of the trough (right half of the central pixel and the redward adjacent 8 pixels) are
$(-0.87\pm0.55)\times 10^{-18}$  and $(1.23\pm0.54)\times 10^{-18}$ erg s$^{-1}$cm$^{-2 }$\AA, respectively.

Without any detailed radiative transfer model or statistical
distribution of profile shapes we can still
obtain a very simple lower limit to the total flux by {\it assuming that
all the flux emerges in the red half of the profile}. This amounts to 
ignoring the blue peak entirely and is probably correct to better than
10\% (judging from the spectra in R08). The integrated
flux in the red half of the trough, covering wavelengths from  [1215.67,1218.2]\AA\ or velocities in the range  from [0, 624] kms$^{-1}$, ,i.e., the wavelength region occupying the RHS half of fig 1, is $(3.11\pm1.38)\times10^{-18}$ erg s$^{-1}$cm$^{-2}$.
This assumes that the zero-level was correctly determined in the
spectrum by Rahmani et al 2010.  However,  the zero level was adjusted globally for the trough and the flux level is sloping so both halves of the trough cannot
be simultaneously at the correct zero level. If we enforce the trivial condition that the flux must be equal or greater than zero in both halves of the DLA trough
simultaneously we are forced to raise the zero level flux density in the entire region at least by the amount of $0.87\times 10^{-18}$erg s$^{-1}$cm$^{-2}$\AA. This leads to zero flux in the blue side, and to a corrected flux density for the red half of $(2.11\pm0.78)\times 10^{-18}$ erg s$^{-1}$cm$^{-2}$\AA. The total flux in the red side, over the interval [0, 624] kms$^{-1}$ is then $(5.35\pm1.97)\times10^{-18}$ erg s$^{-1}$cm$^{-2}$. 

The results are robust with respect to different choices in averaging
the spectra.
Applying the same analysis to the unweighted mean spectrum (Rahmani et al
panel 1 in their fig.1) we get similar numbers, a required upward correction
of the zero level by $0.97\times 10^{-18}$erg s$^{-1}$cm$^{-2}$\AA,
and a  total flux in the red side of $(4.9\pm2.6)\times10^{-18}$ erg s$^{-1}$cm$^{-2}$.

The errors here are purely statistical
and do not take into account the uncertainty regarding the shape of the true flux density distribution.
The flux value given here should systematically underestimate the true flux in several ways: 
We have ignored any flux in the blue peak, have raised the global continuum level only so far that the blue side is zero, and have limited ourselves
to take into account  only flux within [0, 624] kms$^{-1}$ red-ward of the rest frame wavelength. 
If the emission profiles are similar to those of the brighter observed emitters detected  by   R08 
(or those of typical Lyman break galaxies with detectable Ly$\alpha$ emission), these shortcomings may, however, be relatively unimportant at the level of accuracy achieved here.

We emphasize that the above result is only a marginal detection. The method of stacking many low signal-to-noise spectra from the SDSS still falls short  by a factor 4-6 (Rahmani et al. erroneously give a factor 1.7)  of the sensitivity  reached by R08 in detecting individual Ly$\alpha$ emitters. Much  larger samples of DLAS spectra 
(or larger S/N  of individual spectra)  will be required to match
the R08 detection threshold.

Finally, we would like to note that the selection criteria for the 
emitters in R08 and Rahmani et al 2010 are rather  different. The stacking
of DLA troughs  by definition is sensitive to the flux from DLAS host galaxies 
within 1.5" (or at least a significant fraction of the flux for the 
less extended objects).  The emitters of R08 were instead detected when part of their Ly$\alpha$ emission intersected a randomly positioned long slit. HI ionizing photons 
can be efficiently converted into Ly$\alpha$ at column densities below those
characteristic of DLAS (e.g., Gould \& Weinberg 1996) and the Ly$\alpha$ can be scattered into the line-of-sight at even lower column densities (e.g. Barnes \& Haehnelt 2010). Some of the Ly$\alpha$ 
photons received from the R08 emitters should thus originate from regions with  column densities characteristic of sub-DLA or LLS. In consequence, sometimes only a minor fraction of their total luminosity may be recorded by the resulting spectrum. Most of these galaxies are still likely to host DLAS,  but their
luminosities may (in some cases) be severely underestimated by only part of their light falling through the slit. Note further that the redshift distribution of the Rahmani et al. sample 
extends to higher redshift.
A more precise quantitative comparison 
%intending to distinguish between different models for the host  galaxies of DLAS
%based on luminosities 
will need to take these observational differences into account.

\section[]{Conclusions}

Ly$\alpha$ emitters associated with optically thick gaseous halos show asymmetric and mostly redshifted line profiles that, in aggregate, can produce
a tilt of the flux level at the bottom of a DLA absorption profile. We identify such a pattern in the stacked DLAS spectrum presented  by Rahmani et al 2010
and estimate the flux emerging from the trough. Taking into account the expected effects of resonant scattering on the spectral distribution of the emission  and the probable over-subtraction of the sky background at the
bottom of the profile, we arrive at a much higher (but still only marginally significant) estimate for the average Ly$\alpha$ flux than Rahmani et al.
Our lower limit of Ly$\alpha$ flux emerging at the bottom of the stacked DLAS profile (clipped weighted mean) gives $(5.35\pm1.97)\times10^{-18}$ erg s$^{-1}$cm$^{-2}$. 
This value is within one standard deviation of the mean flux of the R08 emitters, $3.7\times10^{-18}$ erg s$^{-1}$cm$^{-2}$, in  reasonable agreement with that 
independent estimate.  The average luminosity of Ly$\alpha$ emitters causing DLAS in the line-of-sight to background QSOs is similar or perhaps even somewhat larger than  the average luminosity of faint Ly$\alpha$ emitters in the R08 sample.  Thus, contrary to the claim by Rahmani et al, our present analysis lends additional support to the conclusions by Rauch et al 2008 that their Ly$\alpha$ emitters largely overlap with the host galaxies of DLAS.

\section*{Acknowledgments}

We acknowledge helpful discussions with George Becker.

\bsp

\label{lastpage}


\begin{thebibliography}{}

\bibitem[Barnes \& Haehnelt 2010]{bar10}Barnes, L. A.,  Haehnelt, M. G. 2010, MNRAS, 403, 870

\bibitem[Djkstra et al 2006]{dij06} Dijkstra, M., Haiman, Z., Spaans, M.,  2006, ApJ, 649, 14

\bibitem[Hansen \& Oh 2006]{han06} Hansen, M., Oh, S.P., 2006, MNRAS, 367, 979

\bibitem[Gould \& Weinberg 1996]{gou96} Gould, A., Weinberg, D.H., 1996, ApJ, 468, 462

\bibitem[Laursen et al 2010]{lau10}Laursen, P., Sommer-Larsen, J., Razoumov, A. O., 2010, astroph/1009.1384

\bibitem[Ouchi et al 2008]{ouc08} 
	Ouchi, M., Shimasaku, K., Akiyama, M., Simpson, C., Saito, T.,  Ueda, Y., Furusawa, H., Sekiguchi, K., Yamada, T., Kodama, T., Kashikawa, N., Okamura, S.,
 Iye, M., Takata, T., Yoshida, Mi., Yoshida, Ma.,  2008 ApJS, 176, 301

\bibitem[Rahmani et al. 2010]{rah10} Rahmani, H., Srianand, R., Noterdaeme, P., Petitjean, P., 2010, MNRAS, in press.

\bibitem[Rauch et al 2008]{rau08} Rauch, M., Haehnelt, M. G., Bunker, A., Becker, G., Marleau, F., Graham, J., Cristiani, S., Jarvis, M.,
Lacey, C., Morris, S., Peroux, C., R\"ottgering, H., Theuns, T., 2008, ApJ, 681, 856


\bibitem[Steidel et al 2010]{stei10} Steidel, C. C., Erb, D. K., Shapley, A. E., Pettini, M., Reddy, N., Bogosavljevic, M., Rudie, G. C., Rakic, O., 2010, ApJ, 717, 289

\bibitem[Verhamme et al 2006]{} Verhamme, A., Schaerer, D., Maselli, A., 2006, A\&A, 460, 397



\bibitem[Zheng \& Miralda-Escud\'e 2002]{zhe02} Zheng, Z., Miralda-Escud\'e, J.,  2002, ApJ, 578,  33

\end{thebibliography}
\end{document}